\documentclass[sigconf]{acmart}

\usepackage[table]{colortbl}
\usepackage[inline]{enumitem}
\usepackage[skip=0pt]{caption}

\usepackage[subtle]{savetrees}

\copyrightyear{2020} 
\acmYear{2020} 
\setcopyright{acmlicensed}\acmConference[SIGIR '20]{Proceedings of the 43rd International ACM SIGIR Conference on Research and Development in Information Retrieval}{July 25--30, 2020}{Virtual Event, China}
\acmBooktitle{Proceedings of the 43rd International ACM SIGIR Conference on Research and Development in Information Retrieval (SIGIR '20), July 25--30, 2020, Virtual Event, China}
\acmPrice{15.00}
\acmDOI{10.1145/3397271.3401297}
\acmISBN{978-1-4503-8016-4/20/07}

\begin{document}
\fancyhead{} 

\title[An Analysis of Mixed Initiative and Collaboration in Information-Seeking Dialogues]{An Analysis of Mixed Initiative and Collaboration\\ in Information-Seeking Dialogues}



  

\author{Svitlana Vakulenko}
\affiliation{%
  \institution{University of Amsterdam}
  \city{Amsterdam} 
  \country{The Netherlands}}
\email{s.vakulenko@uva.nl}

\author{Evangelos Kanoulas}
\affiliation{%
  \institution{University of Amsterdam}
  \city{Amsterdam} 
  \country{The Netherlands}}
\email{e.kanoulas@uva.nl}

\author{Maarten de Rijke}
\affiliation{%
  \institution{\mbox{}\hspace*{-4mm}\mbox{University of Amsterdam \& Ahold Delhaize}}
  \city{Amsterdam} 
  \country{The Netherlands}}
\email{m.derijke@uva.nl}

\renewcommand{\shortauthors}{Vakulenko, Kanoulas and de Rijke}

\begin{abstract}
  The ability to engage in mixed-initiative interaction is one of the core requirements for a conversational search system. 
  How to achieve this is poorly understood.
  We propose a set of unsupervised metrics, termed \emph{ConversationShape}, that highlights the role each of the conversation participants plays by comparing the distribution of vocabulary and utterance types.
  Using ConversationShape as a lens, we take a closer look at several conversational search datasets and compare them with other dialogue datasets to better understand the types of dialogue interaction they represent, either driven by the information seeker or the assistant.
  We discover that deviations from the ConversationShape of a human-human dialogue of the same type is predictive of the quality of a human-machine dialogue.
\end{abstract}






\maketitle


\section{Introduction}
While the idea of conversational search has been around for several decades~\citep[see, e.g.,][]{belkin-1995-cases}, the idea has recently  attracted considerable attention~\citep{radlinski2017theoretical}. 
Conversational user interfaces are believed to facilitate more efficient information access than traditional interfaces.
A conversation, in this case, is a collaborative process that allows an information seeker to satisfy an information need.
One of the key features of a conversational interaction is the potential for \textit{mixed initiative}, where ``the system and user both can take initiative as appropriate''~\cite{radlinski2017theoretical}. 
What is an appropriate action on the part of the system? How initiative can be measured?
In this paper we report on the first attempts to analyse and evaluate the degree of initiative and collaboration between the conversation participants given only conversation transcripts as input.

Conversational search tasks proposed to date mostly reduce dialogues to a sequence of question-answer pairs~\cite{dalton2019trec,radlinski2019coached,AliannejadiSigir19}.
In the datasets used for the question-answering tasks, the structure of interactions is fixed in advance: either the user takes the initiative and the system follows-up with an answer or vice versa, which makes them unsuitable for studying how the initiative is transferred between roles.
Discussion threads from on-line Q\&A forums are a popular data source for developing conversational search tasks~\cite{InforSeek_Response_Ranking,penha2019introducing}.
While on-line forums are a valuable resource for studying real-world interaction patterns, they exhibit a type of asynchronous information exchange, which, as we show in our analysis, is very different from synchronous dialogue interactions.

Conversational systems are often grouped into question answering, task-oriented and chit-chat~\cite{INR-074}.
It is important to note, that this classification schema is mainly based on the difference in approaches to building such conversational systems rather than the difference in dialogues they produce.
In this paper, we focus on analysing and measuring differences between dialogue types, and report on the resulting dimensions and a new dialogue classification scheme that emerges from our analysis.
We show that human-to-human dialogues collected for conversational search tasks bear structural similarities to task-oriented and chit-chat dialogues.



Recent evaluation studies of chit-chat dialogue models show that conversational systems tend to seize control of the dialogue by asking too many questions and ignoring the user's initiative~\cite{DBLP:conf/cui/BowdenWCJHSSWW19,DBLP:journals/corr/abs-1902-00098}.
Standard evaluation metrics do not capture this dimension of a dialogue interaction and, therefore, fail to predict user engagement~\cite{DBLP:journals/corr/abs-1902-00098}.
The most popular metric for dialogue evaluation is relevance of the response, which is usually measured with respect to a ground-truth response; it is comparable to answer accuracy if the response is an answer.
Our work is complementary to this research.
We introduce a novel evaluation framework based on a set of unsupervised features.
The framework is designed to capture the quality of dialogue interactions in terms of balancing initiative, when appropriate, and measuring collaboration between the dialogue participants.


Our evaluation framework is based on several independent lexical features that capture initiative and collaboration in dialogues.
Simple automated measures based on discourse features, such as lexical and syntactic diversity, were previously adopted to reduce repetitive generic responses and to estimate question complexity~\cite{see2019what,DBLP:conf/conversations/JalotaTMNU19}.
We use an unsupervised approach, similar to the ones applied in language style matching~\cite{ireland2010language} and in measuring quality of generated narratives~\cite{DBLP:journals/corr/abs-1909-10705}, to a dialogue setting.
A key characteristic of a dialogue is that it is a type of narrative with utterances generated by multiple dialogue participants.
Therefore, we estimate lexical features separately for each participant so as to be able to compare their contributions and, thereby, deduce the roles they play in the conversation.

The work most similar to ours is by \citet{DBLP:conf/acl/WalkerW90}, who studied lexical cues, such as the use of anaphora and different utterance types, as a mechanism for switching control between dialogue participants.
Our approach to dialogue representation is unsupervised and domain-independent, which allows us to scale an analysis that was previously performed only on a handful of dialogues to thousands of publicly available dialogue transcripts.


Our main contributions can be summarised as follows:
\begin{enumerate*}
\item We examine structural patterns of initiative and collaboration across ten datasets with more than 97k dialogues. Ours is the first study that automatically identifies these dimensions within large and diverse dialogue corpora, drawing parallels between dialogue tasks that originate within different research communities.
\item The initiative and collaboration patterns we identified correlate with human judgements of dialogue quality.
\end{enumerate*}
Allocation of control, where control is defined as managing the direction of flow in a conversation, is at the core of mixed-initiative dialogue systems that are designed to enhance human-machine collaboration~\cite{Cohen1998}.
Dialogue systems should be able to recognize the user’s cues for initiative switch so as to provide appropriate responses~\cite{DBLP:journals/umuai/Chu-CarrollB98}.
Detecting initiative is also important to characterize the quality of the interaction~\cite{Cohen1998}.
Our work contributes insights that inform the design of evaluation and optimisation approaches capable of recognising initiative distribution in dialogue.

\section{ConversationShape}
\label{sec:representation}

\emph{ConversationShape} is a dialogue representation approach that focuses on the structural properties of a dialogue.
We consider a dialogue to be a sequence of utterances exchanged between several participants.
All dialogues in our experiments have two participants.
However, our approach can be also applied to multi-party conversations.
Information-seeking dialogues are often characterised by asymmetry of the roles that participants play in the conversation: one usually assumes the role of an assistant ($A$), whose function is to be automated by a conversational search system; another dialogue participant is an information seeker ($S$), who is using the service of the assistant to obtain information.
To model mixed initiative in dialogues we use four metrics that are calculated separately for each of the dialogue participants: 
\begin{enumerate*}
\item question;
\item information;
\item repetition; and
\item flow.
\end{enumerate*}

\emph{Question} is an explicit attempt at controlling the direction of a conversation flow, since a posed question sets an expectation for another participant to produce a relevant answer.
We trained a supervised classifier on the NPS Chat Corpus~\cite{DBLP:conf/semco/ForsythandM07} to recognize questions and other utterance types.
The NPS Chat Corpus contains 7.9K utterances from online chat rooms, annotated with 14 utterance types: Statement, Emotion, Greet, Bye, Accept, Reject, whQuestion, ynQuestion, yAnswer, nAnswer, Emphasis, Continuer, Clarify, Other.
Our classification model was initialised from a pre-trained RoBERTa~\cite{DBLP:journals/corr/abs-1907-11692} (base model) and further fine-tuned for the utterance type prediction task achieving F1 of 0.81 on the held-out test set.

The remaining metrics describe patterns of collaboration and control over the topic of a conversation.
To explain them, we need to introduce the concept of dialogue vocabulary first.
The \textit{dialogue vocabulary} consists of all unique words (or subword tokens) that occur in the same dialogue transcript.
We are especially interested in the words that occur frequently (more than once) within the same dialogue, since the repetition patterns are likely to signal their importance to the topic of a dialogue.

\emph{Information} reflects the contribution that a participant made to the topic of a conversation.
We estimate information as a count of frequent tokens that were first coined by a conversation participant.

\emph{Repetition} indicates a follow-up on the topic of the conversation.
To analyse the emergence of a shared vocabulary we trace vocabulary reuse patterns between the conversation participants.
We estimate repetition as the number of tokens that were first introduced by one conversation participant and subsequently repeated by another conversation participant.
We consider repetition as a type of relevance feedback available within the conversation, assuming the act of repetition to be an endorsement of the importance of the token to the topic of a conversation by increasing the token frequency.
Another way to reference previous tokens implicitly is to use anaphora.
Therefore, we add the count of anaphora to the count of repetitions.
We use a short list of English anaphora from the analysis framework proposed by \citeauthor{DBLP:conf/acl/WalkerW90}: {`it', `they', `them', `their', `she', `he', `her', `him', `his', `this', `that'}.
We also experimented with an off-the-shelf co-reference resolution model instead but were not satisfied with the results.

\emph{Flow} is the difference between \emph{Repetition} and \emph{Information}, which reflects on the role of a participant in maintaining the coherence of the conversation by referencing previous statements or driving the conversation forward by introducing new information.

For every conversation we compute the values for each of the conversation participants separately: $\mathit{Concept}_A$ and $\mathit{Concept}_S$, where \emph{Concept} denotes one of the four metrics that we have just introduced.
To be able to compare between conversations of different length we also normalise the scores by the number of utterances in a conversation.
Then we use the average and the difference between the two metrics to characterise the type of dialogues in a dataset.
The average shows the magnitude for each of the metrics, e.g. the average number of questions per conversation:
\begin{equation}
    \mathit{Concept} = \frac{\mathit{Concept}_A + \mathit{Concept}_S}{2}.
\end{equation}
The difference allows to compare the distribution (balance) between the dialogue participants, e.g. who asks more questions in a conversation.
We use the formula similar to the one used for the writing style matching~\cite{ireland2010language}:
\begin{equation}
    \Delta \mathit{Concept} = \frac{\mathit{Concept}_A - \mathit{Concept}_S}{\mathit{Concept}_A + \mathit{Concept}_S}.
\end{equation}
It not only indicates the difference in metrics between the roles but also its direction: negative values indicate dominance by the Seeker and positive -- by the Assistant.





\begin{example}
Let us consider a snippet from the dialogue transcript in the Redial dataset~\cite{li2018towards} to illustrate our approach to measuring different aspects of initiative and collaboration in dialogue:

\begin{description}[nosep]
\item[\textbf{(A)}] Hey! What kind of \textit{movies} do you like to watch?
\item[\textbf{(S)}] I'm really big on indie romance and dramas
\item[\textbf{(A)}] Ok what's your favorite \textit{movie}?
\item[\textbf{(A)}] Staying with \textit{that} genre, have you seen @88487 or @104253
\item[\textbf{(A)}] Those are two really good ones
\item[\textbf{(S)}] When I was a kid I liked \textit{horror} like @181097
\item[\textbf{(A)}] @Misery is really creepy but really good. I only recently got into \textit{horror}.
\end{description}

\noindent%
Assistant (A) clearly dominates the conversation asking all the questions ($\mathit{Question_A}=2/7=0.29; \mathit{Question_S}=0; \mathit{Question}=0.15$; $\Delta \mathit{Question}=1$).
A introduced \textit{movies} as the topic of the conversation but subsequently followed up on the topic directions introduced by S: \textit{that} genre and \textit{horror} ($\mathit{Information_A}=\mathit{Information_S}=1/7=0.14; \mathit{Information}=0.14; \Delta \mathit{Information}=0$; $\mathit{Repetition_A}=2/7=0.29; \mathit{Repetition_S}=0; \mathit{Repetition} = 0.29/2=0.15; \Delta \mathit{Repetition}=1$; $\mathit{Flow}_A=\mathit{Repetition_A}-\mathit{Information_A}=0.29-0.14=0.15$; $\mathit{Flow}_S=-0.14$).
We use this approach in the next section to automatically distinguish between dialogues of different types.
\end{example}



\section{Datasets}
\label{sec:datasets}
Our analysis spans across 10 publicly available dialogue datasets, which were designed for various dialogue tasks.
Numbers in brackets indicate the number of dialogues in each of the datasets.

\begin{itemize}[leftmargin=*,nosep]
    \item \textbf{CCPE} (502) -- conversational preference elicitation~\cite{radlinski2019coached}.
    \item \textbf{SCS} (37) -- spoken conversational search~\cite{trippas2018informing}.
    \item \textbf{MSDialog} (35.5K) -- discussion threads from a support forum~\cite{InforSeek_Response_Ranking}.
    \item \textbf{MultiWOZ} (10.4K) -- multi-domain task-oriented dialogues~\cite{DBLP:journals/corr/abs-1810-00278}.
    \item \textbf{ReDial} (10K) -- conversational movie recommendation~\cite{li2018towards}.
    \item \textbf{WoW} (22.3K) -- chatting over topics from Wikipedia~\cite{dinan2018wizard}.
    \item \textbf{DailyDialog} (11K) -- sample dialogues for English learners~\cite{DBLP:journals/corr/abs-1710-03957}.
    \item \textbf{Meena} (91), \textbf{Mitsuku} (100), \textbf{Human} (95) -- human-machine and human-human open-domain dialogues~\cite{DBLP:journals/corr/abs-2001-09977}.
    \item \textbf{ConvAI2} (3.5K), \textbf{Control-M} (3.2K), \textbf{Control-H} (102) -- human-machine and human-human persona-grounded dialogues~\cite{see2019what,DBLP:journals/corr/abs-1902-00098}.
\end{itemize}{}

\section{Results}
\label{sec:results}

\begin{figure}[t]
  \centering
  \includegraphics[width=\linewidth]{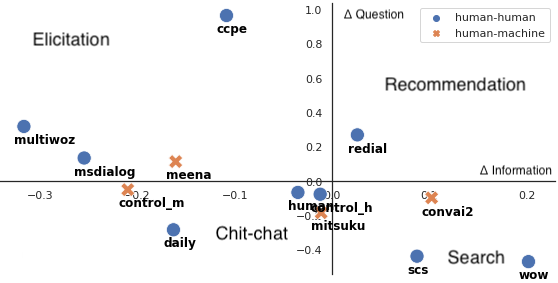}
  \caption{Dialogue types: datasets above the y-axis contain dialogues where questions are mostly posed by the Assistant (Assistant-driven dialogues), below -- by the Seeker (Seeker-driven dialogues); in the dialogues on the right of the x-axis the conversation topic is mainly contributed to by the Assistant, on the left -- by the Seeker.}
 \label{fig:dataset_types}
\end{figure}

Table~\ref{tab:datasets} shows the average \emph{ConversationShape} for each of the dialogue sets from the previous section.
This representation allows to compare the sets and identify different dialogue types, e.g., Figure~\ref{fig:dataset_types} shows the clusters that emerge based on the similarities in Question and Information distributions.

\paragraph{\textbf{Assistant-driven dialogues}}
From Table~\ref{tab:datasets} we see that in CCPE the Assistant  leads the conversation by asking the questions and the Seeker follows up by answering them (negative $\Delta\mathit{Repetition}$).
MultiWOZ and MSDialog also have the majority of questions posed by the Assistant but those questions follow-up on the questions and answers provided by the Seeker (positive $\Delta\mathit{Repetition}$).
In ReDial the Assistant drives the conversation by providing information and asking questions, while the Seeker follows up (negative $\Delta\mathit{Repetition}$).


\paragraph{\textbf{Seeker-driven dialogues}}
SCS and WoW are similar to each other: for both the Seeker is mainly asking questions and the Assistant is providing information.
However, the Seeker follows-up on the topics introduced by the Assistant in WoW (negative $\Delta\mathit{Repetition}$), while in SCS the Assistant follows the Seeker.
Chit-chat dialogues (Human and Control-H) appear closer to the origin showing that the initiative is more balanced between the participants in this dialogue type.
Whereas, in the DailyDialog dataset the initiative is skewed towards the initiator of the conversation, who is more likely to ask questions and set the conversation topic.

\paragraph{\textbf{Model diagnostics}}
ConversationShape can help to evaluate dialogue models and understand the type of deviant behavior the model exhibits.
These experiments were performed on the subsets of the Control-M dataset that correspond to the transcripts produced by different dialogue models.
In total, there are 28 models and we compute a ConversationShape for each of them separately.
Then, we measure cross-entropy between the models' distributions and the distribution calculated for the subset of human-human dialogues (Control-H).
Finally, we compare our results with the human evaluation results reported in the original paper~\cite{see2019what}.
The model with the lowest cross-entropy (0.01) to the human-human distributions was also the model that was preferred by human judges with respect to Interestingness and characterised by better flow and more information sharing.

\begin{figure}[h]
  \centering
  \includegraphics[width=\linewidth]{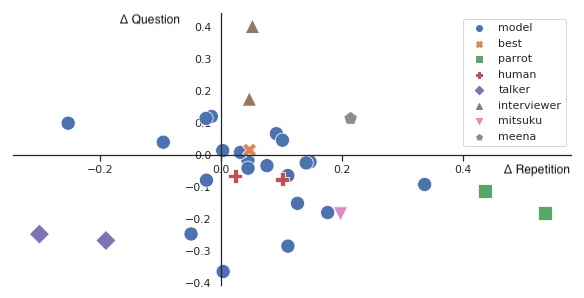}
  \caption{Model diagnostics with all models of the Control-M dataset unrolled. More successful dialogue models tend to ask more questions than in a typical human chit-chat (best and meena). This strategy adaptation does not imply, however, that these models are equally good at following the initiative and answering human questions.}
  \label{fig:model_types}
\end{figure}

Moreover, ConversationShape allows to interpret the type of a deviation a dialogue model exhibit.
We correctly identified the models that either ask too many questions (optimised for inquisitiveness, \textit{interviewer}), repeat too much (optimised for responsiveness, \textit{parrot}) or do not follow up (optimised for diversity or negative responsiveness, \textit{talker}) as outliers in Figure~\ref{fig:model_types}.
We could not achieve the same results when comparing the transcripts of Meena and Mitsuku dialogues~\cite{DBLP:journals/corr/abs-2001-09977}.
The question distribution shows that Meena and Mitsuku dialogues are structurally very different from each other and from typical human chit-chat distribution.
Mitsuku is being interrogated while Meena takes over the initiative by asking questions.

\begin{table*}[]
\caption{ConversationShapes of the popular dialogue datasets listing average and difference in Question, Information and Repetition between the conversation participants. \textbf{Bold font} highlights the highest values for each of the metrics. Grey marker indicates negative values, where the averages are skewed towards the Seeker.}
 \label{tab:datasets}
\begin{tabular}{lcccccccc}
\toprule
Dataset & \emph{Question} & $\Delta$\emph{Question} & \emph{Information} & $\Delta$\emph{Information} & \emph{Repetition} & $\Delta$\emph{Repetition} & \emph{Flow}$_A$ & \emph{Flow}$_S$ \\
\midrule
CCPE~\cite{radlinski2019coached}                     &0.21 (0.03) & \textbf{0.97} (0.10) & 0.23 (0.10) & \cellcolor{lightgray}-0.11 (0.37) & 0.29 (0.09) & \cellcolor{lightgray}-0.27 (0.30) & 0.02 (0.12) & 0.10 (0.16)                       \\
MultiWOZ~\cite{DBLP:journals/corr/abs-1810-00278}    & 0.22 (0.07) & 0.32 (0.40) & 0.39 (0.14) & \cellcolor{lightgray}\textbf{-0.32} (0.39) & 0.44 (0.15) & 0.32 (0.35) & 0.31 (0.31) & \cellcolor{lightgray}-0.22 (0.28)\\
ReDial~\cite{li2018towards}                          & 0.10 (0.05) & 0.27 (0.56) & 0.13 (0.08) & 0.03 (0.58) & 0.21 (0.09) & \cellcolor{lightgray}-0.20 (0.41) & 0.04 (0.15) & 0.12 (0.15)                       \\
MSDialog~\cite{InforSeek_Response_Ranking}           & 0.08 (0.09) & 0.14 (0.68) & \textbf{1.13} (0.71) & \cellcolor{lightgray}-0.25 (0.53) & \textbf{1.02} (0.55) & \textbf{0.46} (0.42) & \textbf{0.64} (1.13) & \cellcolor{lightgray}\textbf{-0.85} (1.07) \\
WoW~\cite{dinan2018wizard}                           & 0.14 (0.08) & \cellcolor{lightgray}-0.47 (0.66) & 0.37 (0.15) & 0.20 (0.50) & 0.52 (0.18) & \cellcolor{lightgray}-0.03 (0.37) & 0.05 (0.38) & 0.24 (0.35)                     \\
SCS~\cite{trippas2018informing}                      & 0.14 (0.06) & \cellcolor{lightgray}-0.44 (0.39) & 0.40 (0.28) & 0.09 (0.39) & 0.45 (0.21) & 0.21 (0.35) & 0.12 (0.40) & \cellcolor{lightgray}-0.02 (0.37) \\
DailyDialog~\cite{DBLP:journals/corr/abs-1710-03957} & 0.17 (0.09) & \cellcolor{lightgray}-0.28 (0.69) & 0.16 (0.15) & \cellcolor{lightgray}-0.16 (0.62) & 0.30 (0.19) & 0.08 (0.53) & 0.20 (0.27) & 0.08 (0.27)                       \\
Control-H~\cite{see2019what}                         & 0.20 (0.08) & \cellcolor{lightgray}-0.08 (0.46) & 0.18 (0.10) & \cellcolor{lightgray}-0.01 (0.58) & 0.22 (0.11) & 0.10 (0.60) & 0.05 (0.18) & 0.02 (0.21)                       \\
Human~\cite{DBLP:journals/corr/abs-2001-09977}                         & 0.21 (0.06) & \cellcolor{lightgray}-0.07 (0.35) & 0.30 (0.13) & \cellcolor{lightgray}-0.04 (0.40) & 0.36 (0.14) & 0.02 (0.33) & 0.08 (0.22) & 0.03 (0.20)                       \\
Meena~\cite{DBLP:journals/corr/abs-2001-09977}                         & 0.20 (0.06) & 0.11 (0.47) & 0.14 (0.06) & \cellcolor{lightgray}-0.16 (0.50) & 0.17 (0.08) & 0.21 (0.42) & 0.08 (0.13) & \cellcolor{lightgray}-0.02 (0.11) \\
Mitsuku~\cite{DBLP:journals/corr/abs-2001-09977}     & 0.20 (0.06) & \cellcolor{lightgray}-0.18 (0.47) & 0.15 (0.10) & \cellcolor{lightgray}-0.01 (0.53) & 0.22 (0.10) & 0.20 (0.39) & 0.11 (0.19) & 0.03 (0.16)  \\
Control-M~\cite{see2019what}                         & \textbf{0.25} (0.07) & \cellcolor{lightgray}-0.05 (0.43) & 0.14 (0.09) & \cellcolor{lightgray}-0.21 (0.65) & 0.16 (0.10) & 0.06 (0.61) & 0.07 (0.19) & \cellcolor{lightgray}-0.03 (0.20) \\
ConvAI2~\cite{DBLP:journals/corr/abs-1902-00098}     & 0.18 (0.14) & \cellcolor{lightgray}-0.10 (0.63) & 0.09 (0.10) & 0.10 (0.50) & 0.09 (0.11) & 0.08 (0.50) & 0.01 (0.19) & \cellcolor{lightgray}-0.00 (0.15)   \\
\bottomrule
\end{tabular}
\end{table*}


\section{Conclusion}
\label{sec:conclusion}
In this paper, we introduced the ConversationShape framework, which provides a set of simple but effective unsupervised metrics designed to measure initiative and flow of a conversation.
Our analysis uncovers relations between different dialogue types and suggests a set of dimensions that are appropriate to consider when developing and evaluating conversational systems, or collecting new dialogue datasets.
Our $\mathit{Repetition}$ metric, which estimates follow-ups on a conversation topic is rather crude since it considers only lexical matches and anaphors.
Though we show that it suffices for a high-level analysis of dataset distributions, predicting quality of the individual dialogues requires a more fine-grained inspection.
Future work should focus on developing an extension that can also account for semantic similarity between tokens.
The next step will be to incorporate these metrics into an optimisation criteria of a learning algorithm that can supply the model with an appropriate perspective on the flow of a conversation and give an explicit incentive to control for an appropriate balance, which, as we showed, depends on the type of the conversation.

\section*{Acknowledgements}
This research was supported by
NWO (016.Vidi.189.039 and 314-99-301),
H2020 (814961), and
Google Faculty Research Award.
All content represents the opinion of the authors, which is not necessarily shared or endorsed by their respective employers and/or sponsors.

\balance
\bibliographystyle{ACM-Reference-Format}
\bibliography{refs}

\end{document}